# Causal explanation for observed superluminal behavior of microwave propagation in free space

W. A. Rodrigues, Jr[(1,2,*)], D. S. Thober[(1)], A. L. Xavier, Jr[(1)]

(1)  *Wernher von Braun-Advanced Research Center, UNISAL,*
*Av. A. Garret, 267, 13087-290 Campinas, SP Brazil*
(2) *Institute of Mathematics, Statistics and Scientific Computation, IMECC-UNICAMP*
*CP 6065, 13083-970, Campinas, SP Brazil*

*In this paper we present a theoretical analysis of an experiment by Mugnai and collaborators where superluminal behavior was observed in the propagation of microwaves. We suggest that what was observed can be well approximated by the motion of a superluminal X wave. Furthermore the experimental results are also explained by the so called scissor effect which occurs with the convergence of pairs of signals coming from opposite points of an annular region of the mirror and forming an interference peak on the intersection axis traveling at superluminal speed. We clarify some misunderstandings concerning this kind of electromagnetic wave propagation in vacuum.*

In this note we make some comments on a recent paper by Mugnai, Ranfagni and Ruggieri [1] claiming the observation of superluminal behavior in microwave propagation for long distances in air. Our comments are of theoretical nature and about possible explanations for the measured effect.

From the theoretical point of view, we start by analyzing the following statement quoted in [1] and attributed to [2]: "Yet, there is no formal proof, based on Maxwell equations that no electromagnetic wave packet can travel faster than the speed of light." First of all we note that every physical wave (satisfying Maxwell equations) produced by a physical device of finite dimension (antenna) must necessarily have a beginning (possibly) an end in time, say at $t=-T$ and $t=0$. We say that such an electromagnetic field configuration is an electromagnetic *pulse* of compact support in the time domain. Observe also that if the pulse generated by the device does not spread with an infinity speed, then when it is ready, let us say at $t=0$ it must occupy, due to the finite dimension of the antenna, a finite region in space. Such a pulse is necessarily of *finite energy* [3]. In an appropriate reference frame, we can then write that at $t=0$, the signal has support only for $|\vec{x}| \leq R$, where $R$ is the maximum linear dimension involved. Maxwell equations (in vacuum) are a hyperbolic system of partial differential equations [4]. Moreover, each one of the components of the free electromagnetic field solves a homogeneous wave equation. It is then possible to prove under very general conditions (strictly hyperbolic Cauchy problem) that, if an electromagnetic pulse has field components and first time derivatives with compact support in space at $t=0$, then the time evolution of such field components must be *null* [5] for $|\vec{x}| \geq R + ct$. As usual $c$ is the parameter that appears in the homogeneous wave equation satisfied by any of the components of the electromagnetic field. This result can be called *finite propagation speed* theorem. We emphasize here that this theorem implies that the front of the pulse travels with *maximum* speed $c$ (in some cases we can prove that it indeed travels with speed $c$) but it does not fix any *minimum* speed for the lateral boundary of the signal spread. This is a very important result since it enables the project of antennas for sending well focused waves. However, it is important to stress that perfect focusing is impossible for any finite energy solution of Maxwell equations [6]. Note also, that it is not possible to prove an analogous of the finite propagation speed theorem for waves that do not have compact support in the space domain.

When Maxwell equations are applied to the description of wave motion in dispersive media with losses or gains and under special conditions, the propagation of finite energy electromagnetic pulses (as defined above) may exhibit superluminal (or even negative) group velocities. We recall that Sommerfeld and Brillouin, showed long ago (see [7]) that a plane wave electromagnetic pulse travels with front velocity $c$ even in dispersive media with loss. Sommerfeld and Brillouin result is a very particular one since their plane wave pulse has *infinite energy*. Indeed, Zhou [8] recently found an example where plane wave pulses can have superluminal front velocities when traveling in a special dispersive media with gain (on this issue, take into account also the comment in [3]). Sommerfeld and Brillouin concluded that the concept of group velocity can not be applied when the group velocity becomes superluminal and this statement has been repeated in many textbooks since then, as, e.g., in [9,10]. However, their conclusion is misleading since it is now well known that superluminal group velocities can be observed (see, e.g., [11,12,13]). Even negative group velocities have been observed [14,15]. The explanation for some of these superluminal (or negative) group velocities observed in dispersive media and also in microwave tunneling is found in the *reshaping phenomenon* [16,17]. However there are some claims that this is not the case [18]. The basic argument in favor of *genuine* superluminality, [18], is that a real wave packet must have compact support in the frequency domain because: "...signals with an infinite spectrum are impossible, since Planck has shown in 1900 that the minimum energy of a frequency component is $\hbar\omega$''. We do not intend to discuss this specific argument here. However, we remark that even if we do not take into account the fact that real signals must have as discussed above finite energy, we must have in mind the following fact. Fourier theory implies that a signal with compact support in the frequency domain is unlimited in the time domain, i.e., has no fronts and as such it is *impossible* to define a front velocity for it and only the group velocity has physical meaning. Following this reasoning, we cannot endorse the point of view of [18] which seems also the one adopted in [1], simply because it implies the existence of wave packets in the time range $-\infty < t < \infty$, i.e., even before the antenna was turned on. The concepts presented above, although of fundamental character, are unfortunately not well known as they should be.

We now state a result that at first sight seems to contradict what has been said above, that is: Maxwell equations (and also all the other linear relativistic wave equations) possess exact arbitrary speed solutions ($0 \leq v < \infty$) that are undistorted progressive waves (*UPWs*) even in free space (for a review, see [16] and also [19,20,21]). These *UPWs* solutions, like plane wave solutions of Maxwell equations, have infinite energy, and classical electromagnetic theory implies that they are the only convenient approximations to waves that can be really built in the physical world. There exists therefore a *crucial* distinction between solutions of Maxwell equations and physical realizable solutions of that equations. Once an exact *UPW* solution is known, it is possible to launch pulses that are *finite aperture approximations* to that *UPW* i.e., pulses obtained through the *Rayleigh-Sommerfeld approximation* [21]. Such finite aperture approximations always have fronts that propagate with velocity $c$ and, of course, have *finite energy*. As already stated, there are subluminal, luminal and superluminal *UPWs*. A finite aperture approximation to a subluminal (*superluminal*) *UPW* pulse can be shown theoretically to have a peak travelling at subluminal (*superluminal*) group velocity even if the front travels with velocity $c$!. This phenomenon was predicted and observed for the first time in experiments with acoustic waves [17], where sub and superluminal means $v < c_s$ or $v > c_s$ respectively ($c_s$ the sound speed appearing in the corresponding homogeneous wave equation). In [17] it was predicted that the phenomenon could be observed for electromagnetic superluminal *X* waves. In fact Saari and Reivelt produced a finite aperture approximation to a superluminal *X* wave pulse in the optical range [22].

For clarity and in order to not give chance for any misconception and misunderstanding let us emphasize the following. All of the theoretical (analytical and numerical simulations) studies of real cases of finite aperture approximations to exact superluminal *UPWs* have shown that their peaks indeed travel at $v \geq c$ in some particular circumstances. Since the initial front of any given pulse travels at velocity $c$ it is reached by the peak after some propagation time. This happens when the pulse loses its lateral confinement after a distance called the depth of the field and starts to behave like an ordinary spherical wave. This may be called a *generalized reshaping phenomenon* whose origin is obvious [16]. An immediate consequence of the generalized reshaping phenomenon is the qualitative prediction [16] that the velocity of the peak must decrease along the propagation direction. This phenomenon is exactly what has been observed in the experiment reported in [1].

After these necessary preliminaries, we can now present specific criticisms to the contents of [1]. Recall that we can generate solutions of Maxwell equations with the Hertz potential method [9,16]. The Hertz potential satisfies a homogeneous wave equation for a free Maxwell equation. If we take the form used in [17], e.g., a magnetic Hertz potential $\Pi_m = \Phi \hat{z}$ in a fixed direction, say the $z$-direction taken as the propagation direction, then $\Phi$ satisfies a homogeneous wave equation, namely

$$\frac{\partial^2 \Phi}{c^2 \partial t^2} - \nabla^2 \Phi = 0. \qquad (1)$$

The most simple and non-trivial solution of this equation in cylindrical coordinates $(\rho, \varphi, z)$ is

$$\Phi(t, \rho, z) = J_0(\Omega \rho) \exp\left[-i(\omega t - k_z z)\right], \qquad (2)$$

$$\Omega^2 = (\omega/c)^2 - k_z^2. \qquad (3)$$

The solution given by eq.(2) has cylindrical symmetry, i.e., it is independent of the $\varphi$ variable. In Eq.(2) $J_0$ is the zeroth order Bessel function and $\Omega$ is a separation constant (see, e.g., [19,17] for details). eq.(3) representing a dispersion relation at first sight looks strange, but it is rigorously true (details of this derivation can be found in [19]).

It is very important to emphasize that eq.(3) does not imply that the $J_0$ wave function will propagate with *distortion* in vacuum. The crucial meaning of that dispersion relation is that implies that the Hertz potential associated with the field given by eq.(2) and its associated electromagnetic fields are *not* superluminal. Only the phase velocity is superluminal, the group velocity remains subluminal. This interpretation, as showed in [17] is indeed correct, since it is possible to find a Lorentz reference frame where the solution represents a standing wave. Also, for the acoustical case, as reported in [17], a *Bessel beam*, i.e., a finite aperture approximation to the wave packet of the form

$$\Phi_B = J_0(\Omega \rho) F^{-1}[T(\omega)] \exp(ik_z z), \qquad (4)$$

where $T(\omega)$ is an appropriate transfer function and $F^{-1}$ is the inverse Fourier transform, is such that its peak was reported to travel at subluminal speed (i.e., with $v < c_s$). This eventually surprising result is rigorously proved in [19].

Another important fact to be clarified is that *no* electric component of the Maxwell field can have the same form as the Hertz potential in eq.(2). As can be easily verified by direct computation, an electric Hertz potential, $\Pi_e = \Phi \hat{z}$ (see [23]) with $\Phi$ as in eq.(2) naturally generates besides a transverse component that is given by a term proportional to a $J_1$ function, also a *longitudinal* electric field composed of three terms, one of which with a $J_0$ dependence. The explicit formulas for the fields are given in [19]. We note moreover that we succeeded in deriving analytically the form of the diffracted electric field produced by a ring aperture as happens in the case of the experiment reported in [1]. Only for a particular mathematical model of the transfer function of the ring (a delta function), the diffracted electric field has a transverse component which is a $J_0$ function multiplied by a spherical outgoing wave and a phase factor, but a longitudinal component is present.. If this transverse component passes through a focusing lens under appropriate conditions it results in a transverse electric field that is a $J_0$ function. A realistic transfer function gives a more complicated field. These results will be reported elsewhere. In conclusion, we must say that the experiment reported in [1] that refers to a horn antenna emitting a TE wave *cannot* be represent by the field given by Eq. (2).

So, in conclusion we can say no finite aperture approximation to an electromagnetic Bessel beam of the form of Eq.(4) (which includes eq.(2) as a particular case) can show any superluminality. As a consequence, if we accept the data in [1] as correct, we must conclude that no finite aperture approximation to an electromagnetic $J_0$ *Bessel beam* was observed.

But if this is the case, what kind of wave could produce the superluminal effect described in [1]? A proposed answer is that what have been observed is a particular kind of a *finite aperture approximation* to a superluminal electromagnetic *X* wave pulse. Acoustical *X* pulses were firstly produced by Lu and Greenleaf in the [24,25]. The speeds of the acoustical *X* pulses have been measured in an experiment reported in [18], where also the mathematical theory of superluminal *X* waves (and their finite aperture approximations) and computer simulations for their behavior were presented. A particularly simple superluminal *X* wave can be generated through the magnetic Hertz potential $\Pi_m = \Phi_X \hat{z}$, where $\Phi_X$ is a packet of Bessel waves of the form given by eq.(2). Putting

$$\Omega = k \operatorname{sen}\theta, \quad k_z = k \cos\theta, \qquad (5)$$

we have a new dispersion relation,
$$\omega/c = k \qquad (6)$$

but, of course, and this is crucial, the propagation vector continues to be $k_z$. We can now verify that

$$\Phi_X(t,\rho,z) = \int_{-\infty}^{\infty} dk B(k) J_0(k\rho \operatorname{sen}\theta) e^{-ik(z\cos\theta - ct)}$$
(7)

is a superluminal wave as solution of the homogeneous wave equation. In eq.(7) $B(k)$ is a frequency distribution function and $\theta$ is called the *axicon* angle. Note that we obtained $\omega/c = k$, at the cost of making the separation parameter frequency dependent, something that is really an extraordinary idea and apparently has been first introduced by Fujiwara [26] and after that used by Durnin [27,28] in pionner papers on well focused waves. Theoretically, the waves represented by eq.(7) are *UPWs* and move with *genuine* superluminal speed $v = c/\cos\theta$ whichever be the frequency distribution $B(k) \neq \delta(k - k_0)$. When $B(k) = \delta(k - k_0)$ we are back to eq.(2). Now, we already explained that in [1] it was *not* the function given by eq.(2) that represents the electromagnetic *pulse* of the experiment because it has infinite energy. Also, the field configuration given by (2) has support for $-\infty < t < \infty$ and as such cannot correspond to the waves that the authors used in their experiment, for they emphasized that they launched electromagnetic pulses, i.e., wave packets with compact support in the time domain. Then, even if it is possible (see below) to find frequencies distributions such that eq.(7) represent pulses (i.e., waves of compact support in the time domain), these pulses cannot be the electromagnetic pulses used in the experiment because they also have infinite energy.

Then, the question is: how to *model* the launching of a pulse? The modelling must be given in two steps. First we must mathematically model in a correct way a superluminal solution of Maxwell equations that can be associated to the problem and after that we must project finite aperture approximations to that solution. Here, we examine for simplicity, the case of a pulse representing a magnetic Hertz potential as above, launched from a plane antenna located on the $z = 0$ plane and such that it solves the homogeneous wave equation together with appropriated boundary conditions (Sommerfeld conditions) at $z = 0$ [29]. For our problem the appropriated boundary conditions that produce a solution like the one given by eq.(7) and represents moreover a superluminal $J_0$ pulse (a *X* wave) must be (see appendix)

$$\bar{\Phi}_X(t,\rho,0) = J_0(\rho\omega_0 \sin\theta) e^{-i\omega_0 t} \mathrm{T}(t),$$

$$\frac{\partial \bar{\Phi}_X(t,\rho,0)}{\partial z} = i\omega_0 \cos\theta J_0(\rho\omega_0 \sin\theta) e^{-i\omega_0 t} \mathrm{T}(t),$$
(8)

with $\mathrm{T}(t) = [\Theta(t) - \Theta(t-T)]$ and $\Theta(t)$ being the Heaviside function) is such that it propagates rigidly (i.e., without distortion) with superluminal speed. Indeed, in this case the function $B(k)$ in eq.(7) is given by eq.(A7) of the appendix and the integral in eq.(7) gives

$$\Phi_X(t,\rho,z)$$
$$= \begin{cases} J_0(\rho\omega_0 \sin\theta)e^{-i\omega_0(t-\cos\theta z)}, & |t-z\cos\theta| < T, \\ 0, & |t-z\cos\theta| > T. \end{cases}$$
(9)

We see then that the peak as well the front of the pulse propagates with the same speed $c/\cos\theta$! Of course, this result *cannot* be taken as a proof that there exists a physically realizable wave whose front propagates in vacuum with superluminal velocity. This is so because the solution just found has *infinite* energy.

Thus, to complete the modeling of the experiment we recall that the real wave (more precisely, the Hertz potential) used in [1] (or in any real experiment) must be represented by a *finite aperture approximation* to the wave $\bar{\Phi}_X$, denoted $\bar{\Phi}_{XFAAA}$, which as already said above has *finite* energy, has a peak which moves with superluminal speed $v = c/\cos\theta$, and a front which moves with speed *c*. Of course, a phenomenon like that cannot last forever and must disappear when the peak catches the front that is traveling at the speed of light, a result predicted in [16] where this effect has been called generalized reshaping phenomenon.

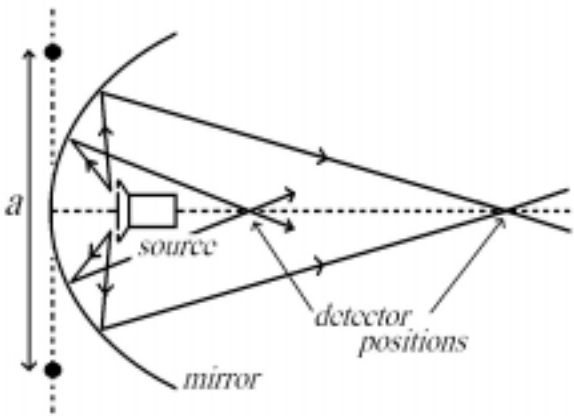

Figure 1: Schematic representation of the launcher (horn antenna) and mirror. The two point source model for the scissor effect is seen on the left separated by the distance $a$.

For a quantitative description of the experiment in [1], it is necessary, of course, to produce a detailed simulation under the correct experimental conditions, i.e., to reproduce a finite aperture approximation to the pulse generated by the ring aperture on the mouth of the horn antenna and which is reflected by the spherical mirror (see Figure 1). This complete simulation will be reported elsewhere.

However, to understand the mechanism behind the superluminality (and the data) observed in the experiment a simple model suffices. Before presenting our simple model we discuss an important issue raised by the authors of [1], i.e., we show that a purely *geometrical* description of the rays in the apparatus cannot reproduce the superluminal velocities reported in [1]. A simulation using geometric optics and naturally including the spherical aberration of the mirror is shown in Fig. 2 for the two axicon angles $\theta = 16^0$ and $\theta = 23^0$ used in the experiment. This simulation estimates the travel time of geometric rays emanating from the feed in Fig. 1 that falls on the mirror and are reflected onto the symmetry axis of Fig. 1 at each specific detector position.

Mugnai, Ranfagni and Ruggieri "signal velocity" for each point along the symmetry z-axis is then determined as follows: consider an annular source (feed) located on the mirror focal plane projecting rays onto the mirror. In Fig.1 the annular source is represented by two point sources, the annular slit as seen edged-on. Reflected rays cross the z-axis at different points depending on the aperture angle of the source rays (axicon angle). Each ray takes a specific time to travel from the reflection point on the mirror surface to the crossing point on the z-axis. Detectors are placed at different positions (distant *L* from each other) on the z-axis (not between mirror and source) and we simply calculate the time difference *T* between the rays reaching these different detectors. The signal velocity is then given by the derivative of the curve *L-T*. The axicon angle obtained by adjusting the diameter of the circular slit changes the pattern of time distribution along the z-axis, the larger the angle the more pronounced the superluminal effect.

We see that for the two axicon angles $\theta = 16^0$ and $\theta = 23^0$ used in the experiment we have an increase of about 4% and 8% for the signal velocity, respectively, in complete disagreement with the values reported in [1] and showing that geometric optics is hardly an explanation for the phenomenon. Indeed, the experimental results exceed such numbers mainly for $\theta = 23^0$ and we agree with authors of [1] that diffraction effects are the cause for the observed velocity in this case. Our statement comes from an accurate simulation of wave propagation in the experiment including diffraction due to source shadowing and will be reported elsewhere.

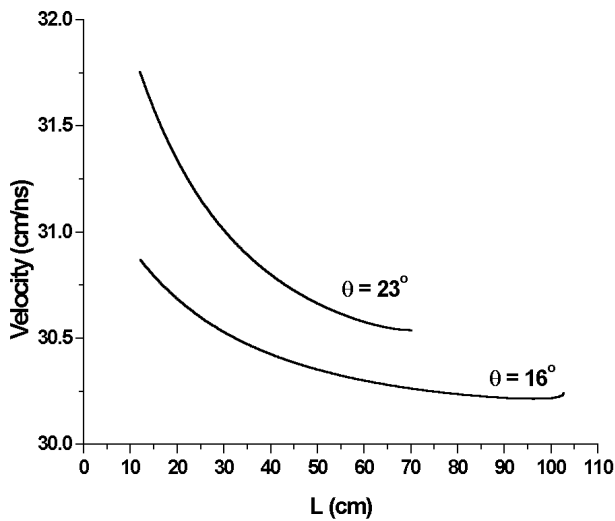

Figure 2: Signal velocity as a function of the detector position estimated using geometrical ray tracing. The velocity profile does not agree with the values found in [1].

Having showed that simple geometric optics cannot explain the experimental data of the experiment we now introduce our promised model. It is an *approximate model* that explains the mechanism behind the superluminal velocities observed in the experiment and fits the measured effect for $\theta = 16^0$ and $\theta = 23^0$. We already said that a single $X$ pulse given by eq.(9) cannot explain the experiment in [1] because it propagates rigidly at *constant* superluminal speed and the data reported in [1] shows a *varying propagation* speed $v(z)$ of the peak. In order to proceed, we recall that for $|t - z\cos\theta| > T$ a X-wave pulse like the one given by eq.(9) can be written as

$$J_0(\rho\omega_0 \sin\theta)e^{-i\omega_0(t-\cos\theta z)}$$
$$= e^{-i\omega_0 t}\int_0^{2\pi} d\varphi e^{\{ik\sin\theta[x\cos\varphi + y\sin\varphi] + ik\cos\theta z\}} \quad (10)$$

i.e., an integral over the polar angle $\varphi$ (in the $z = 0$ plane) of plane wave pairs emitted from points of a circle in the plane with angles $\varphi$ and $\pi - \varphi$ and traveling at speed *c*. This suggests to think of the real wave in the experiment as an interference pattern generated by sequence of spherical waves emanating from simultaneous sources (an annular region) on the mirror, which move with velocity *c* and interfering on the z-axis. With this supposition the dependence of $v$ on $z$ can be quantitatively explained. The resulting wave fronts propagate in the $z$ direction (and are in fact *perpendicular* to it) but, in essence, each of its *components* propagates at a tilted angle in relation to the *z*-axis. This is indeed the base of the so called *scissor effect*. The reader should understand that the purpose of our simple model is to illustrate a possible mechanism that produces the phenomenon and not to give accurate values for the measured velocities. For simplicity we admit a "virtual" annular source with diameter *a* placed somewhere on a plane behind (or in front of) the mirror. If $t$ is the time counted since the production of the spherical pulse, then the scissor speed on the z-axis is given by (as it is easy to verify)

$$v(t) = \frac{ct}{\sqrt{t^2 - a^2/4c^2}}, \quad (11)$$

and therefore the distance covered by the main scissor peak along *z*-axis until time *t* is

$$L(t_z, t_1) = L_0 + \int_{t_1}^{t_z - T_d} v(t)dt, \quad (12)$$

where $t_1 \geq a/2c$ is some reference time, $L_0$ is an off-set length and $T_d$ is a time delay. We can therefore fit a curve to the experimental results in [1] based on three parameters: An offset in *z*, $L_0$, an offset in time, $T_d$, and the distance between the sources (diameter of the annular source), *a*. Another parameter would be the position along *z* of the virtual source, but for simplicity, we choose this position at $z=0$. Since the interference peak is on the z-axis, it is only locally similar to a X-wave. There is no simple relation between the axicon angle of the local $X$ pulse - which changes along $z$ - and the axicon angle used in [1] on the assumption that it was indeed a Bessel pulse (as imagined by the authors of [1]). A larger circular slit radius (for the real source on the mirror focal plane) simply implies a larger separation of the virtual sources in our model.

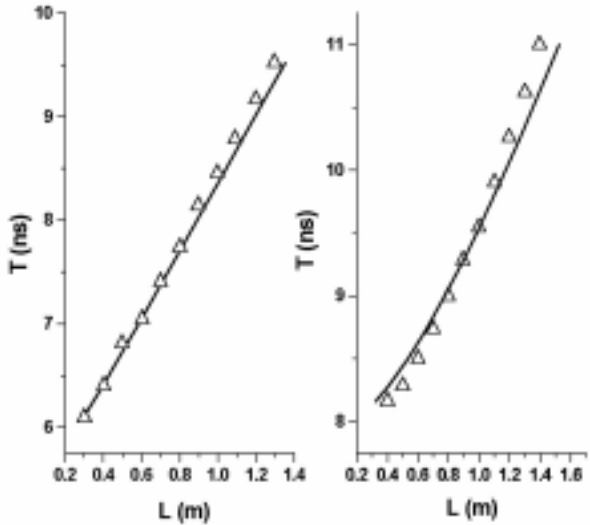

Figure 3: Fit of the delay time measurements as a function of distance $L$ along the z-axis using the scissor effect model. Triangles represent measured data extracted from [1]. Left: $\theta = 16^0$, Right: $\theta = 23^0$.

Numerical simulations are shown in Figure 3 and, and given the simplicity of the model and the inexistence of error bars in the original data of [1], they agree reasonably well. For the case $\theta = 16^0$ we used $L_0 = 0$, $T_d = 5.1$ $ns$ and $a = 10\,cm$. For $\theta = 23^0$ the fitting parameters were $L_0 = -2\,cm$, $T_d = 5.0$ $ns$ and $a = 182\,cm$.

We conclude the paper stressing that the experiment in [1] shows explicitly that a kind of generalized *reshaping phenomenon* occurs under appropriate conditions even for pulses propagating in free space (in the case of [1] we recall that air, the medium where the propagation occurs, is transparent to microwaves). The origin of the generalized reshaping is crystal clear from our discussion [16]. There is no question of principle involved in the experiment. The finite aperture approximations to superluminal $X$ waves produced by the experiment are of compact support in the time domain and of finite energy and their fronts propagate always with the speed $c$. Only the peaks of these pulses travel at superluminal speed and they are detected in the experiment instead of the fronts due to the limited detection threshold of the receivers. The peak of any pulse however does not causally connect source to detector, leaving relativity theory intact. Also, the phenomenon of superluminal motion cannot last indefinitely. In fact it lasts until the peak catches the front, defining the maximum distance (called field depth of order $(a/2)\cot\theta$, see [28]) for which a finite aperture approximation to a superluminal $X$ wave is reasonably focused. A simple explanation for the superluminal motion reported in [1] is given by the interference of spherical wave fronts on the symmetry axis. The interference pattern builds the superluminal peak and constitutes the well known scissor effect.

## Acknowledgements

Authors are grateful to Motorola do Brasil for financial support (Project *X*) and ALXJr would like to thank financial support by FAPESP (Fundação de Amparo à Pesquisa do Estado de São Paulo) under contract number 00/031680. Also thanks are due to Professor E. Capelas de Oliveira for discussions and to a referee whose comments helped to improve the quality of the paper.

## Appendix

Here we show why the boundary conditions given by eq.(8) must be used in order to have a solution of the homogeneous wave equation of the form given eq.(7). We also determine for these conditions the function $B(k)$. The mathematical formulation of our problem is as follows: find a solution of the homogeneous wave equation with cylindrical symmetry (i.e., independent on the variable $\varphi$ such that at the $z=0$ plane we have the following boundary conditions,

$$F_1(t,\rho) = \Phi(t,\rho, z=0),$$
$$F_2(t,\rho) = \left.\frac{\partial \Phi(t,\rho,z)}{\partial z}\right|_{z=0}. \quad (A1)$$

As it is well known, the general solution can be obtained from a wave packet of functions of the from given by eq.(2). In this appendix for simplicity we use units such that $c=1$. We write [9]

$$\Phi(t,\rho,z) = \frac{1}{4\pi}\int_{-\infty}^{\infty} d\omega A(\omega,\rho)e^{ik_z(\omega)z}e^{-i\omega t}$$
$$+ \frac{1}{4\pi}\int_{-\infty}^{\infty} d\omega S(\omega,\rho)e^{-ik_z(\omega)z}e^{-i\omega t}. \quad (A2)$$

From eq.(3) we have $k_z(\omega) = \sqrt{\omega^2 - \Omega^2}$. We introduce now a frequency dependent separation constant $\Omega$, writing $k_z = k\cos\theta$, $\Omega = k\sin\theta$. This implies $\omega = k$ and we rewrite eq.(A3) as

$$\Phi(t,\rho,z) = \frac{1}{4\pi}\int_{-\infty}^{\infty} d\omega A(\omega,\rho)e^{i\omega\cos\theta z}e^{-i\omega t}$$
$$+ \frac{1}{4\pi}\int_{-\infty}^{\infty} d\omega S(\omega,\rho)e^{-i\omega\cos\theta z}e^{-i\omega t}. \quad (A3)$$

Then,
$$A(\omega,\rho) = \frac{1}{4\pi}\int_{-\infty}^{\infty} dt\left[F_1(t,\rho) - i\frac{F_2(t,\rho)}{\omega\cos\theta}\right]e^{i\omega t},$$
$$S(\omega,\rho) = \frac{1}{4\pi}\int_{-\infty}^{\infty} dt\left[F_1(t,\rho) + i\frac{F_2(t,\rho)}{\omega\cos\theta}\right]e^{i\omega t}. \quad (A4)$$

We want a solution without the $S(\omega,\rho)$ term. Then we must have
$$F_2(t,\rho) = i\omega\cos\theta F_1(t,\rho) = ik_z(\omega)F_1(t,\rho). \quad (A5)$$

From eq.(A5) we see that if $F_1(t,\rho)$ is given by the first of the eqs.(8) then $F_2(t,\rho)$ must be given by the second of eqs.(8).

The function $B(k)$ in eq.(7) can now be found using the following identity [19] valid for $T = 2\pi n/\omega_0$ with $n$ an integer,

$$\frac{1}{2\pi}\int_{-\infty}^{\infty} d\omega\left[\frac{e^{i(\omega-\omega_0)T}-1}{i(\omega-\omega_0)}\right]J_0(\rho\omega\sin\theta)e^{-i\omega(t-\cos\theta z)}$$
$$= \begin{cases} J_0(\rho\omega_0\sin\theta)e^{-i\omega_0(t-\cos\theta z)} & \text{for } |t-z\cos\theta| < T, \\ 0 & \text{for } |t-z\cos\theta| > T. \end{cases}$$
(A6)

Therefore it follows that
$$B(k) = \frac{1}{2\pi}\left[\frac{e^{i(\omega-\omega_0)T}-1}{\omega-\omega_0}\right]. \quad (A7)$$

## References and notes

[1] D. Mugnai, A. Ranfagni and R. Ruggieri, *Phys. Rev. Lett.* **84**, 4830 (2000).
[2] S. Bosanac, *Phys. Rev. A* **28**, 577 (1983).
[3] This observation is of fundamental importance, since it has been proved in [19] that there exist pulses of finite time duration but of infinite energies such that their fronts propagate at superluminal speed even in the vacuum!
[4] R. Courant and D. Hilbert, *Methods of Mathematical Physics*, vol. 1 (J. Wiley and Sons, New York, 1996).
[5] M. E. Taylor, *Pseudo Differential Operators* (Princeton Univ. Press, Princeton, 1981).
[6] T. T. Wu and H. Lehmann, *J. Appl. Phys.* **58**, 2064 (1985).
[7] L. Brillouin, *Wave Propagation and Group Velocity* (Academic Press, New York, 1960).
[8] X. J. Zhou, *Phys. Lett* **A** 278, 1 (2001).
[9] J. A. Stratton, *Electromagnetic Theory* (McGraw-Hill, New York, 1941).
[10] J. D. Jackson, *Classical Electrodynamics*, third edition (J. Wiley and Sons, New York, 1999).
[11] A. M. Steinberg, P. G. Kwiat and R. Y. Chiao, *Phys. Rev. Lett.* **71**, 708 (1993).
[12] E. L. Bolda, J. C. Garrison and R. Y. Chiao, *Phys. Rev. A* **49**, 3938 (1994).
[13] L. J. Wang, A. Kumzmich and A. Dogariu, *Nature* **406**, 277 (2000).
[14] C. G. B. Garret and D. E. McCumber, *Phys. Rev. A* **1**, 305 (1970).
[15] T. Emig, *Phys. Rev. E* **54**, 5780 (1996).
[16] J. E. Maiorino and W. A. Rodrigues, Jr., *What is Superluminal Wave Motion?*, electronic book (at http\\:www.cptec.br, *Sci. and Tech. Mag.* **4**, 1999)
[17] W. A. Rodrigues, Jr. and J. Y. Lu, *Found. Phys.* **27**, 435 (1997).

[18] G. Nimtz, *Europhys. J. B* **7**, 523 (1999).
[19] E. C. Oliveira, W. A. Rodrigues,Jr., A. L. Xavier and D. S. Thober, "Thoughtful Comments on Bessel Beams and Signal Propagation", to appear in *Phys. Lett A* (2001).
[20] E. C. de Oliveira and W. A. Rodrigues, Jr., *Ann. Phys.* (Berlin) **7**, 654 (1998).
[21] P. M. Morse and H. Feshback, *Methods of Theoretical Physics*, Part 1 (McGraw-Hill, New York, 1953).
[22] P. Saari and K. Reivelt, *Phys. Rev. Lett*. **79**, 4135 (1997).
[23] A. Enders and G. Nimtz, *Phys. Rev. B* **47**, 9605 (1993)
[24] J. Y. Lu and J. L. Greenleaf, *IEEE Transact. Ultrason. Ferroelec. Freq. Cont*. **39,** 19 (1992).
[25] J. Y. Lu and J. L. Greenleaf, *IEEE Transact. Ultrason. Ferroelec. Freq. Cont*. **39,** 441 (1992).
[26] S. Fujiwara, *J. Opt. Soc. Am*. **52**, 287 (1962).
[27] J. Durnin, *J. Opt. Soc. Am. A* **4**, 651 (1987).
[28] J. Durnin, J. J. Miceli, Jr. and J. H. Eberly, *Phys. Rev. Lett.* **58** (1987) 1499.
[29] The case of the pulse reflected by a parabolic antenna as used in [1] will be reported elsewhere.